# Nonlinear dynamical systems and linearly forced isotropic turbulence

Zheng Ran

Shanghai Institute of Applied Mathematics and Mechanics,
Shanghai University, Shanghai 200072, P.R. China

In this letter, we present an extensive study of linearly forced isotropic turbulence. By using an analytical method, we identified two parametric choices that are new to our knowledge. We proved that the underlying nonlinear dynamical system for linearly forced isotropic turbulence is the general case of a cubic Lienard equation with linear damping (Dumortier and Rousseau 1990).

**PACS** number: 47.27.Gs, 47.27.ed, 47.20.Ky

The possibility of a dynamical system approach allows one to capture fundamental physical mechanisms such as the energy cascade in 3D turbulence[1]. In particular, one can build a bridge between the traditional description in statistical terms and the dynamical behaviour in phase space. For fluids under severe constraints, i.e. the experiments on Rayleigh-Bénard convection in small cells, the success was undisputed[2]. For understanding fully developed isotropic turbulence, however, success has been more limited. We observe that in the present stage of development of the theory of dynamical systems, there has been little quantitative effect on the understanding of a high Reynolds number flow[3].

Here, we outline a new method to obtain a natural second-order nonlinear dynamical system based on the exact solution of the linearly forced Karman–Howarth equation. Maintaining a turbulent flow in a more or less stationary state, for better statistics in experiment or for convenience in theoretical considerations, requires forcing the flow. Various types of forcing schemes have been developed. A special forcing scheme was proposed [4] by which we may simplify the deterministic models to the bare minimum, in some sense, assuming that the usually velocity-dependent force term is merely proportional to the velocity field at all positions at all times. The linear forcing scheme was further studied by several groups [5,6]. We will consider this type of forced isotropic turbulence governed by the incompressible Navier–Stokes equations:

$$\frac{\partial \vec{u}}{\partial t} + (\vec{u} \cdot \nabla)\vec{u} = -\frac{1}{\rho}\nabla p + \nu \nabla^2 \vec{u} + \phi \vec{u}, \quad (1)$$

where the incompressibility condition reads

$$\nabla \cdot \vec{u} = 0, \quad (2)$$

and $\phi$ is a positive constant with dimensions of inverse time.

The two-point double and triple longitudinal velocity correlation, denoted by $f(r,t)$ and $h(r,t)$, respectively, are defined in a standard way[7]. The Karman–Howarth equation derived from it under the conditions of homogeneity and isotropy reads

$$\frac{\partial}{\partial t}(bf) + 2b^{\frac{3}{2}}\left(\frac{\partial h}{\partial r} + \frac{4h}{r}\right) = 2vb\left(\frac{\partial^2 f}{\partial r^2} + \frac{4}{r}\frac{\partial f}{\partial r}\right) + 2\phi bf, \qquad (3)$$

where $(r,t)$ are the spatial and time coordinates, $v$ is the kinematics viscosity, and $b = <u^2>$ denotes the turbulence intensity. Following von Karman and Howarth, we introduce the new variables

$$\xi = \frac{r}{l(t)}, \qquad (4)$$

where $l = l(t)$ is a uniquely specified similarity length scale. As already noted in the case of free decay isotropic turbulence[8,9,10,11], equation (3) could be reduced to the following systems:

[1] The two-point double and triple longitudinal velocity correlation, denoted by $f(r,t)$ may be written as

$$\frac{d^2 f}{d\xi^2} + \left(\frac{4}{\xi} + \frac{a_1}{2}\xi\right)\frac{df}{d\xi} + \frac{a_2}{2}f = 0, \qquad (5)$$

with boundary conditions $f(0) = 1$, $f(\infty) = 0$, and $a_1$ and $a_2$ are constant coefficients.

[2] The turbulent scale equations read

$$\frac{dl}{dt} = a_1 \cdot \frac{v}{l} + 2I_1 \cdot \sqrt{b}, \qquad (6)$$

$$\frac{db}{dt} = -a_2 \cdot \frac{vb}{l^2} + 2\phi b + 2I_2 \cdot \frac{b^{\frac{3}{2}}}{l}, \qquad (7)$$

where $I_1$ and $I_2$ are the two constants of integration.

[3] For the third-order correlation coefficient, from the system of Karman–Howarth equations, one can derive the following equation:

$$\frac{dh}{d\xi} + \frac{4}{\xi}h = -\frac{l}{2b^{\frac{3}{2}}}\frac{db}{dt}f + \frac{1}{2\sqrt{b}}\frac{dl}{dt}\xi\frac{df}{d\xi} + \frac{v}{\sqrt{b}l}\left(\frac{d^2 f}{d\xi^2} + \frac{4}{\xi}\frac{df}{d\xi}\right) + \frac{\phi l}{\sqrt{b}}f. \qquad (8)$$

In fact, the problem is thus reduced to a solvable system; furthermore, one can prove that there is a self-closed second-order nonlinear dynamical system for $l = l(t)$:

$$\frac{d^2 l}{dt^2} = -(2a_1 + a_2)\cdot\frac{v}{2l^2}\cdot\frac{dl}{dt} + (a_1 a_2 v^2)\cdot\frac{1}{2l^3} + \phi\cdot\frac{dl}{dt} - \phi\cdot(a_1 v)\cdot\frac{1}{l} + \frac{I_2}{2I_1}\cdot\frac{1}{l}\cdot\left\{\frac{dl}{dt}\right\}^2$$
$$-(a_1 v)\cdot\frac{I_2}{I_1}\cdot\frac{1}{l^2}\cdot\frac{dl}{dt} + (a_1 v)^2\cdot\frac{I_2}{2I_1}\cdot\frac{1}{l^3}. \qquad (9)$$

Let

$$z = l^{-2}, \qquad (10)$$

or

$$l = z^{-\frac{1}{2}}.  \tag{11}$$

The scale equation for $z = z(t)$ is

$$\frac{d^2z}{dt^2} - \left\{\frac{3}{2} - \frac{I_2}{4I_1}\right\} z^{-1} \cdot \left[\frac{dz}{dt}\right]^2 = -(2a_1 + a_2) \cdot \frac{v}{2} \cdot z \cdot \left\{\frac{dz}{dt}\right\} - \left(a_1 a_2 v^2\right) \cdot z^3 +$$
$$+ \phi \cdot \left\{\frac{dz}{dt}\right\} + 2\phi \cdot (a_1 v) \cdot z^2$$
$$- (a_1 v) \cdot \frac{I_2}{I_1} \cdot z \cdot \left\{\frac{dz}{dt}\right\}  \tag{12}$$
$$- (a_1 v)^2 \cdot \frac{I_2}{I_1} \cdot z^3.$$

In order to avoid the new freedom, the forcing condition is given by

$$\frac{3}{2} - \frac{I_2}{4I_1} = 0.  \tag{13}$$

Using this forcing condition, equation (12) can be reduced to the following form

$$\frac{d^2z}{dt^2} + [-\phi + Az] \cdot \frac{dz}{dt} + Bz^3 + Cz^2 = 0,  \tag{14}$$

where

$$A = \frac{(2a_1 + a_2)v}{2} + (a_1 v) \cdot \frac{I_2}{I_1},  \tag{15}$$

$$B = \left(a_1 a_2 v^2\right) + (a_1 v)^2 \cdot \frac{I_2}{I_1},  \tag{16}$$

$$C = -2\phi \cdot (a_1 v).  \tag{17}$$

In order to use the standard mathematical results for a nonlinear dynamical system[11], the transformation is performed by the formula

$$z = ax + c,  \tag{18}$$

where $a$ and $c$ are different constants. In this case, we obtain

$$a\frac{d^2x}{dt^2} + \{-\phi + A \cdot [ax + c]\} \cdot \left[a \cdot \frac{dx}{dt}\right] + B \cdot \left\{a^3 x^3 + 3a^2 cx^2 + 3ac^2 x + c^3\right\} +$$
$$+ C \cdot \left\{a^2 x^2 + 2acx + c^2\right\} = 0.  \tag{19}$$

The transformation conditions can be written in the explicit form

[1] The coefficient of the term $x^3$ is equal to 1:

$$B \cdot a^2 = 1.  \tag{20}$$

[2] The coefficient of the term $x^2$ is zero:

$$a \cdot (3cB + C) = 0. \tag{21}$$

Consequently, we have

$$a = \pm \frac{1}{\sqrt{B}}, \tag{22}$$

$$c = -\frac{C}{3B}. \tag{23}$$

The corresponding equation can be presented in the form

$$\frac{d^2x}{dt^2} + \{(cA - \phi) + (aA)x\} \cdot \frac{dx}{dt} + x^3 + \{3c^2B + 2cC\} \cdot x + \frac{c^3B + c^2C}{a} = 0. \tag{24}$$

It can be easily proven that for these transformations, in spite of their nonlinear dynamical character, a new nonlinear dynamical system may be found in a standard manner as

$$\frac{dx}{dt} = y, \tag{25}$$

$$\frac{dy}{dt} = -x^3 + \mu_2 x + \mu_1 + y \cdot [\bar{b}x + \bar{v}]. \tag{26}$$

Here,

$$\mu_2 = -3c^2B - 2cC, \tag{27}$$

$$\mu_1 = -\frac{c^2}{a} \cdot [cB + C], \tag{28}$$

$$\bar{b} = -(aA), \tag{29}$$

$$\bar{v} = \phi - cA. \tag{30}$$

The above system thus depends on the three constant parameters $(a_1, \sigma, v)$, where $a_1 > 0$, and (Ran 2008, 2009):

$$\sigma = \frac{a_2}{2a_1}. \tag{31}$$

Consequently, the parameters can be found to be

$$A = (a_1 v) \cdot [7 + \sigma], \tag{32}$$

$$B = 2(a_1 v)^2 \cdot [3 + \sigma], \tag{33}$$

$$C = -2\phi \cdot (a_1 v). \tag{34}$$

As we find from the definition

$$\bar{b} > 0. \tag{35}$$

Therefore,
$$A > 0. \tag{36}$$

The physical intepretation of $a$ is
$$a = -\frac{1}{\sqrt{B}} < 0. \tag{37}$$

As a result, we find
$$z = -\frac{1}{(a_1 v)\sqrt{2(3+\sigma)}}\left\{x + \frac{\sqrt{2}}{3\sqrt{3+\sigma}}\phi\right\}, \tag{38}$$

$$\mu_2 = \frac{2}{3} \cdot \frac{\phi^2}{3+\sigma}, \tag{39}$$

$$\mu_1 = \frac{4\sqrt{2}}{27} \cdot \frac{\phi^3}{[3+\sigma]^{\frac{3}{2}}}, \tag{40}$$

$$\bar{v} = \phi\left\{1 + \frac{7+\sigma}{3(3+\sigma)}\right\}, \tag{41}$$

$$\bar{b} = \frac{7+\sigma}{\sqrt{2(3+\sigma)}}. \tag{42}$$

Our main results can be summarized as follows.

Using an analytical method, we identify two parametric choices that are new to our knowledge. The underlying nonlinear dynamical system for linearly forced isotropic turbulence is the general case of a cubic Lienard equation with linear damping (Dumortier and Rousseau 1990):

$$\frac{dx}{dt} = y,$$

$$\frac{dy}{dt} = -x^3 + \mu_2 x + \mu_1 + y \cdot [\bar{b}x + \bar{v}],$$

where

$$\mu_2 = \frac{2}{3} \cdot \frac{\phi^2}{3+\sigma},$$

$$\mu_1 = \frac{4\sqrt{2}}{27} \cdot \frac{\phi^3}{[3+\sigma]^{\frac{3}{2}}},$$

$$\bar{b} = \frac{7+\sigma}{\sqrt{2(3+\sigma)}},$$

$$\bar{v} = \phi\left\{1 + \frac{7+\sigma}{3(3+\sigma)}\right\},$$

and where $(\mu_2, \mu_1, \bar{\nu})$ forms the parametric space for linearly forced isotropic turbulence. One can easily check that $\bar{b} > 2\sqrt{2}$.

**Acknowledgements:**
The work was supported by the National Natural Science Foundation of China (Grant Nos. 11172162, 10572083).